\documentclass{PoS}

\PoS{PoS(BDMH2004)075}
\usepackage{oldgerm}

\usepackage{graphicx}
\usepackage{amssymb}

\title{Simulating galaxy clusters : the ICM and the galaxy populations}

\ShortTitle{Simulating galaxy clusters: ICM and galaxies}

\author{\speaker{Alessio D. Romeo}\\ %\thanks{A footnote may follow.}\\
        Univ. di Catania, Italy and TAC Copenhagen, Denmark\\
        E-mail: \email{aro@ct.astro.it}}

\author{{Jesper Sommer-Larsen}\\
        TAC, Nordita and NBI, Copenhagen, Denmark\\
        E-mail: \email{jslarsen@tac.dk}}

\author{{Laura Portinari}\\
        TAC Copenhagen, Denmark and Tuorla Obs, Finland\\
        E-mail: \email{lporti@utu.fi}}

\abstract{Cosmological LambdaCDM TreeSPH simulations of the formation and evolution
of galaxy groups and clusters have been performed. The simulations include:
star formation, chemical evolution with non-instantaneous recycling, metal
dependent radiative cooling, strong star burst and (optionally) AGN driven
galactic super winds, effects of a meta-galactic UV field and thermal
conduction.
We report results on the temperature and entropy profiles of the ICM, the
X-ray luminosity, cold fraction, M/L and IMLR ratios in gas and stars, metal
abundances and gradients. Besides, the properties of the galaxy
populations in the two richest clusters are discussed: global star
formation rates of the cluster galaxies, the total K-band luminosity, the
galaxy luminosity functions at z=0 and their redshift evolution, the
colour-magnitude relation
(``red sequence'') as resulting from metallicity effects, and the role of
the IMF in reproducing colours and abundances of the stellar populations.
Finally the contribute (20-40\%) to cluster light from the intra-cluster
stars and the cD galaxy has been investigated: surface brightness
profiles, mean colours and chemical abundances, kinematics (velocity
dispersion and distributions) of the IC stellar populations.}

\FullConference{Baryons in Dark Matter Halos\\
                 5-9 October 2004\\
                 Novigrad, Croatia}

\newcommand{\gsim}{\raisebox{-3.8pt}{$\;\stackrel{\textstyle >}{\sim}\;$}}
\newcommand{\lsim}{\raisebox{-3.8pt}{$\;\stackrel{\textstyle <}{\sim}\;$}}
\newcommand{\ga}{\gsim}
\newcommand{\la}{\lsim}
\def\etal{{\it et al.~}}
\def\aap{A\&A}

\begin{document}

\section{Introduction}

%The basic scenario describing the thermal behaviour of the ICM calls
%for a gravity-only driven heating of the gas by adiabatic compression,
%followed by hydrodynamical shocks arisen during the accretion of infalling 
%material.
%It is by now well known how this simple model basically fails in reproducing
%the actual scaling properties observed ($L_X-T$, $S-T$, $M-T$ relations)
%Physical mechanisms able to explain the break of self--similarity in the
%properties of ICM have been extensively proposed during last ten years, and
%mainly fall in the two broad classes of non--gravitational heating and
%radiative cooling. 
%However, all cluster simulations with cooling and star formation
%reveal a significant drawback: cooling alone makes a large fraction (30 up
%to 55\%) of gas get converted into a cold phase (either cold gas or stars)
%within the virial radius at $z=0$, in contradiction with a value less than %10-20\% of the baryon fraction locked into stars indicated by observations. 
Non--gravitational heating and radiative cooling have been extensively 
recoursed to in the literature in order to complete the basic gravity-only driven scenario describing the thermal behaviour of the ICM, which 
is unable to reproduce
the scaling relations observed between $L_X$, $T$ and $S$.

So far few hydrodynamical simulations have been performed which were able
to pursue at the same time and in a self--consistent way the target of
realistically modelling the ICM dynamics by including both non-gravitational
heating/cooling (complete with feedback/star formation) {\it and} metal
enrichment directly springing from chemically following up the stellar feedback itself 
%Borgani \etal 2001 and 2002, Tornatore \etal 2003, McCarthy 
%\etal 2004 , and 
(e.g. Tornatore \etal 2004).
Moreover, only recently
%, due to advances in computing capabilities
%as well as in the detail in the physical modelling, cluster
simulations have reached a level of sophistication adequate to even trace
star formation and related effects in individual galaxies, coupled with 
the chemical enrichment of the ICM by galactic winds.

In this series of papers, we present 
%for the first time (to our knowledge) 
an analysis
of the properties of the galaxy population of clusters as predicted directly
from cosmological simulations that follow detailed baryonic 
physics, gas dynamics and galaxy formation and evolution. The resolution
for such N--body + hydrodynamical simulations cannot reach the extent of  
resolving galaxies at the faint end of the luminosity function 
($M_B$$\ga$--16). On the other hand, our simulations aim at
describing in a self--consistent way the
hydrodynamical response of the ICM to star formation, stellar
feedback and chemical enrichment. 

%\section{The Code and Simulations}

Our cosmological N-body simulation has been performed using the FLY code 
(Antonuccio \etal, 2003) on a standard
$\Lambda$CDM model 
%($\Omega_0$=0.3, $\Omega_{\Lambda}$=0.7, $\Omega_b$=0.04, $h=0.7$, %$\sigma_8$=0.9) with $128^3$ collisionless DM-only particles filling 
%an initial box of 150$h^{-1}$ Mpc at an initial redshift $z_i=39$. 
The selected clusters and groups 
%(listed in Tab. 1) 
isolated from the cosmological simulation,
as identified at $z=0$, have 
been resampled and resimulated (adding the baryon particles according to
a $f_b=0.12$) by means of an improved version of the Tree-SPH 
code described in Sommer-Larsen, G\"otz \& Portinari (2003; SLGP). 
%All DM particles within
%a virial radius at $z=0$ have been traced back to the initial condition at %$z_{i}$,
%when one SPH particle has been added per each DM one, with a mass of $m_{SPH}=
%f_b\cdot m_{DM}^0$ (with $f_b=0.12$).
The mass resolution reached inside the resampled
lagrangian subvolumes from the cosmological set was such that 
$m_{DM}=1.8\cdot 10^9 h^{-1} M_{\odot}$
and $m_{g}=m_*=2.5\cdot 10^8 h^{-1} M_{\odot}$.

Most of the results hereby presented refer to the two larger systems:
``sub-Coma'' ($M_{vir}=12.4\cdot 10^{14}M_{\odot}$, $<kT_{ew}>\simeq$ 6~keV) and
``Virgo'' ($2.8\cdot 10^{14}M_{\odot}$, 3~keV), though four more smaller
clusters and groups have been simulated which span the range 1-2~keV.

The adopted TreeSPH code schematically includes the following features:
conservative solution of entropy equation;
metal-depending atomic radiative cooling;
star formation, according to either an Arimoto-Yoshii (AY) or a Salpeter IMF;
feedback as starburst (SN II) driven galactic winds;
chemical evolution with non-instantaneous recycling of gas and heavy
elements (H, He, C, N, O, Mg, Si, S, Ca, Fe);
thermal conduction;
meta-galactic, redshift-dependent UV field.
All the runs implement the ``Super-Wind" (SW) scheme, characterized by
high efficiency parameters $\epsilon_{SF}$ and $f_{wind}$: the latter
is amplified by a factor 2 or 4 in those runs simulating the effect of
stronger non-stellar feedback (AGN). On the top of this model, it has
been added either thermal conduction (at $\frac{1}{3}$ Spitzer value (COND
runs), or pre-heating in form of 0.75/1.50/50cm$^2$~keV injected per
particle at $z$=0 (PH runs). Finally, we ran as a test the WFB simulation,
which adopts a normal feedback scheme with only early winds at low values of
$f_{wind}$ (as described in SLGP), and the ADIABATIC run without cooling nor
star formation at all. 
%\begin{table}[t]
%\caption{Structural properties of selected clusters at $z$=0
%\label{t:1}}
%\small
%\begin{tabular}{lcccc}
%\hline\hline
%Cluster \# & $M_{tot}$ & $M_{vir}$ & $R_{vir}$ & $<kT_{ew}>$ $^{\rm a}$ \\
%%$^{\rm c}$ & $M_{b,tot}$ $^{\rm d}$ & $N_*$ & $N_{cg}$ $^{\rm b}$ & $N_{hg}$ %$^{\rm c}$ 
%%& $N_{b,tot}$ $^{\rm d}$\\
%  & [10$^{14}$M$_{\odot}$] & [10$^{14}$M$_{\odot}$] & [Mpc] &[keV]~\\
%\hline
%``Coma" & ~~14.0 &12.38 & 2.90 & 6.00\\
%``Virgo" & ~~3.05 & 2.77 & 1.62 & 3.06\\
%202 & ~~1.26 & 1.26 & 1.29 & 2.24\\
%215 & ~~1.17 & 1.03 & 1.18 & 1.48\\
%550 & ~~0.51 & 0.48 & 0.96 & 1.00\\
%563 & ~~0.50 & 0.49 & 0.94 & 1.10\\
%\hline \hline
%\multicolumn{5}{l}{$^{\rm a}$ Temperatures referred to the "standard" run %SW-AY}\\ 
%\end{tabular}
%\end{table}

\section{Results on the ICM}
\subsection{Thermal properties}

\begin{figure}
\centering \includegraphics[width=70mm]{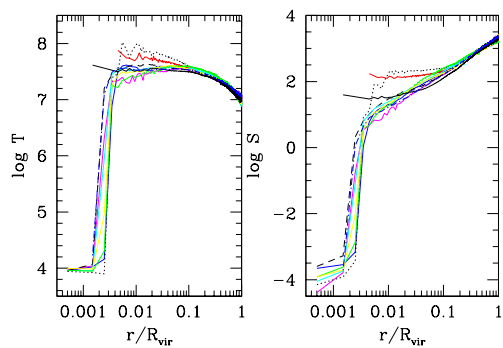}
%  \vspace*{174pt}
  \caption{Temperature ({\it left}) and Entropy ({\it right}) profiles of 
``Virgo'', all models: same colours as in  Fig.~3.}
\label{TSr}               
\end{figure}

Observations of X-ray surface brightness profiles in poor clusters and 
groups of galaxies had hinted at the existence of isentropic cores or 
``entropy floor'' 
at around 100 keV$\cdot$cm$^2$ (e.g. Ponman \etal, 2003), associated with a
reduced gas central density. Systems at all scales, and in particular low-mass ones, show evidence for
excess entropy compared to that expected from pure gravitational heating by compression.
%However, the existence of isentropic cores is still controversial.
In Fig.~\ref{TSr} the inner pattern of temperature and entropy for particles in ``Virgo'' cluster is shown: 
%from which it is evident that there is not a proper isentropic core or entropy %floor: 
most models exhibit central entropy values of
$\sim$ 10-20 keV$\cdot$cm$^2$, which may get agreed with inferred values
in cold--core clusters; 
%(Voit \etal 2003); 
instead a plateau has formed only in two distinct cases, namely the WFB metal-independent scheme with Salpeter IMF and the 4-times SW scheme with AY IMF. The former has less efficient cooling and therefore lets more high-entropy gas settle down in a core.

 The SWx4 model ends up with forming a floor around 100 keV$\cdot$cm$^2$ at $z=0$. 

%\subsection{X-ray Luminosity}

%Measurements of ICM temperature by X-ray
%satellites demonstrated that the observed systems scatter around the curve %$L_X\propto T^\alpha$ with $\alpha \simeq 3$, the largest deviations from which
%are presented by those systems whose emission is strongly associated with a
%cold core, or former ``cooling--flow" clusters. Moreover, there is evidence 
%that low-mass systems ($T\lsim 1$ keV) may
%show an even steeper relation (Helsdon \& Ponman 2000).

\begin{figure}
\centering \includegraphics[width=70mm]{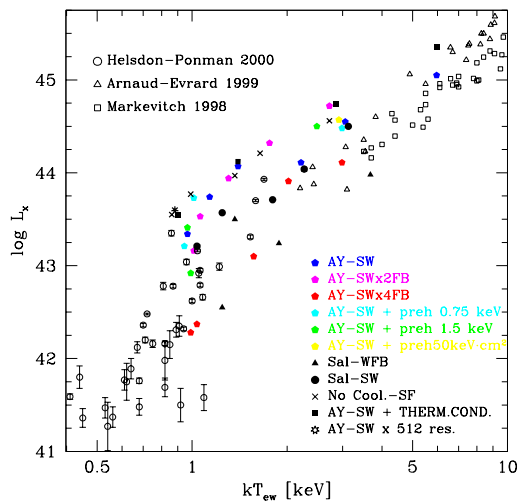}
%  \vspace*{174pt}
  \caption{X-ray luminosity - emission weighted temperature scaling 
relation from the inner 1 Mpc.}
\label{LT}
\end{figure}

In Fig. \ref{LT} the relation $L_X$-$T$ at $z\simeq 0$ is shown: 
%where the temperature is 
%the emission-weighted one over the whole cluster central emitting 
%region (1 Mpc). 
the observed (open) points exhibit the characteristic double-slope
behaviour; as expected, the adiabatic runs lie above this line.
Overall, the predicted luminosities are higher than the observed one; both an enhanced
feedback mechanism and a top--lighter IMF, however, can indipendently concur in 
lowering the X-ray emission from the ICM: the latter effect seems to be dominant,
since the runs with Salpeter IMF and Weak Feedback, which result in a low
level of metal enrichment as well, are those placed lowest. All in all, when strengthening the feedback up to 4 times and using a top--heavy
IMF, a satisfactory normalization seems to be reached. 
%at least for moderate to rich clusters. 
One could conclude then that
previous simulations successful at reproducing both
the L-T relation and the surface brightness profiles and the rather high
entropy floor as well, might have underestimated the contribution of metals to
radiative cooling by adopting simple primordial cooling functions.

\begin{figure}
\centering \includegraphics[width=70mm]{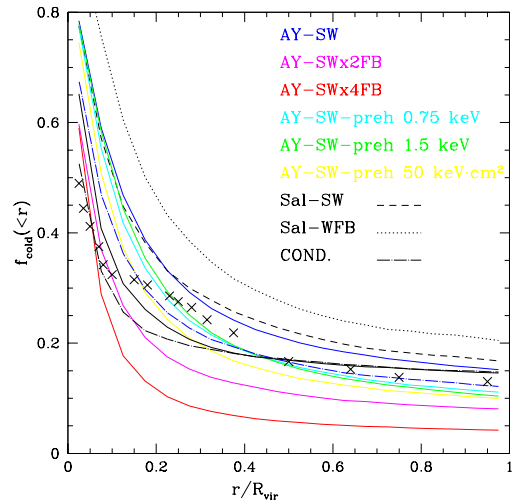}
%  \vspace*{174pt}
  \caption{Cold gas plus stellar radial profiles of the two most massive clusters 
(``Coma'': {\it dot-dashed} line). {\it Crosses}: observed cold fraction 
profiles, derived from the $\frac{M_*}{M_{ICM}}$ data of Roussel {\it et al.} 
(2000).}
\label{FcR}
\end{figure}

%\subsection{The cluster cold component}
Baryonic mass of clusters is largely dominated by the hot ICM mass,
yet it is the (stellar) mass in galaxies playing the major role
to the purpose of chemical enrichment of ICM itself.
A measure of galaxy formation efficiency is
the ``cold fraction'', defined as the fraction of baryons in the form
of cold gas and stars. 
%The cold fraction is a crucial constraint for cluster simulations, as 
It plays
a major role in assessing the mechanisms responsible for the ``excess
entropy'': the larger the cold fraction (hence star formation),
%the larger the role played by galaxy and star formation which removes 
the more low entropy gas is removed from
the core; with a small cold fraction, instead, strong (pre)heating or
feedback are needed to enhance the entropy level.
The radial profiles of the cold fraction are shown in Fig.~\ref{FcR}  
for our two largest simulated clusters:
%, compared to the observed profiles of Roussel {\it et al.} (2000). 
the AY-SW simulations (including the conduction
and preheating cases) are those that best reproduce the overall level, or equivalently the $\frac{M_{ICM}}{L}$ ratio, within the virial radius.
The Salpeter runs instead, 
especially the weak feedback case, result in too large $f_c$, while
the enhanced feedback cases predict significantly too low values.
It is a remarkable result that our simulations with SN feedback 
(with a top-heavy IMF, but without invoking additional sources of energy input 
via enhanced feedback) 
fairly succeed in reproducing the low fraction of baryons in the cold phase. Indeed it is a common feature of all hydrodynamical simulations with 
cooling and star formation ending up with a too large fraction of gas 
to get converted into stars (30 up
to 55\%, against observed 10-20\%). 
When including also non--gravitational (pre)-heating, 
star formation gets attenuated by preventing gas from cooling at high
redshift, hence causing a decrease in the cold fraction.

\subsection{Metal enrichment}

The presence of heavy elements in the hot intra-cluster gas calls 
for the processing of a significant fraction of the ICM inside galaxies. 
%with a tight link between the stellar and gaseous components of clusters.
In our simulations, the super--winds connected to the SNII feedback 
prescription lead to the enrichment of the ICM by 
dispersal of the chemical elements produced by the star particles.
\begin{figure}
\centering \includegraphics[width=70mm]{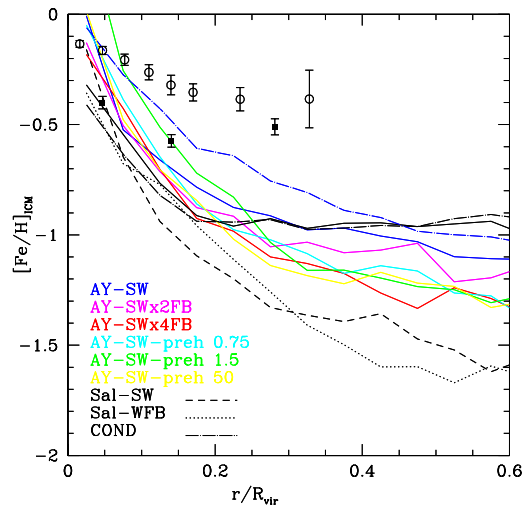}
%  \vspace*{174pt}
  \caption{Iron gradients in the ICM of the two most massive simulated
clusters. Data from De Grandi {\it et al} (2004) refer to cold-core 
({\it open circles}) 
and non cold-core ({\it filled squares}) clusters.}
\label{FehVC}
\end{figure}
Fig.~\ref{FehVC} shows the simulated iron abundance profiles, 
compared
to data for cool core and non-cold core clusters. Our simulated clusters
are best compared to cold-core clusters. Evidently, the simulated
gradients are steeper than the observed ones. This seems to be a general
problem in simulations, 
%(e.g.\ Tornatore {\it et~al.}\ 2004), 
probably related
to the fact that star formation is very concentrated toward
the inner region of the cD galaxy. 
%Another important constraint on the chemical evolution of clusters is the
The ICM is also enriched by $\alpha$--elements, mainly produced by
SNII and specifically traced by silicon abundances: 
We found a trend of supersolar
[Si/Fe] increasing with radius in all our simulations. 

%\begin{figure}
%\centering \includegraphics[width=70mm]{evolICM.eps}
%%  \vspace*{174pt}
%\caption{Radial profiles of Iron ({\it top}) and Silicon ({\it bottom}) 
%abundances in the ICM at z=0,1,2. {\it Left}: ``Virgo'' ({\it solid}: AY-SW, 
%{\it dashed}: Sal-SW); {\it right}: ``Coma''. $R_{vir}$ evaluated at z=0.}
%\label{SiFez}
%\end{figure}

%Fig.~\ref{SiFez} (top panels) shows that the overall iron abundance, and 
Moreover, it has also been found that the overall iron abundance in the ICM 
of simulated clusters is unvaried from $z=0$
to $z=1$, in very good agreement with observations. 
%(Tozzi {\it et~al.}\ 2003).
Since the bulk of the stars in the cluster are formed at $z \geq$2, 
at $z$=2 we predict much lower iron metallicities,
but such redshift range is not probed by observations yet.
%The lower panels in the figure show the evolution in 
As expected, [Si/Fe]
decreases in time due to the delayed iron contribution of SNIa 
with respect to the bulk of the silicon production (from SNII).

\begin{figure}
\centering \includegraphics[width=70mm]{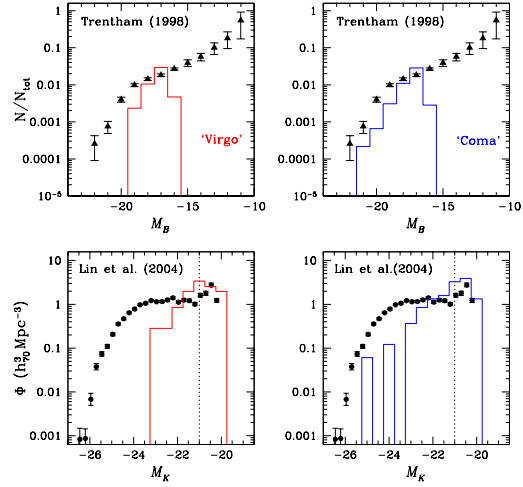}
%  \vspace*{174pt}
  \caption{{\it Top panels}: B--band luminosity functions at $z$=0
for the standard (AY-SW) ``Virgo'' ({\it left}) and ``Coma'' ({\it right})
simulations, compared with the observed composite LF of Trentham 1998, {\it triangles}. 
{\it Bottom panels}:
K--band LFs for the same clusters, compared to the observed average LF by
Lin {\it et~al.}\ (2004); the dotted line marks the limit $M_K \leq -21$ where 
the observational estimate is considered reliable.}
\label{fig:LF_rel}
\end{figure}
%%%%%%%%%%%%%%%%%%%%%%%%%%%%%%%%%%%%%%%%%%%%%%%%%%
\section{Results on cluster galaxies}

\subsection{Luminosity Functions}
\label{sect:lf}
Each star particle represents a Single Stellar Population (SSP), 
with individual stellar masses distributed
according to a particular IMF. We 
keep record of the age and the metallicity 
of each of these SSPs and hence 
the global luminosities and colours of the member galaxies are computed.
%, as the sum
%of the contribution of their constituent star particles. 
We identify for the
standard (SW-AY) runs 42 galaxies in ``Virgo'' and 212 galaxies in ``Coma''.

In Fig.~\ref{fig:LF_rel} 
we compare the B--band and K--band luminosity function (LF) 
of our simulated cluster galaxies, exluding the cD, 
to two observational composite LFs. 
%by Trentham (1998), and by Lin {\it et al.} (2004). 
The number of resolved individual galaxies 
quickly drops for objects fainter than $M_B \sim$ -17, both in ``Virgo'' and in
``Coma'', due to the resolution limits.
Although we are missing the dwarf galaxies that largely dominate
in number, we are able to describe the bulk of the stellar mass and of the
luminosity in clusters, which is dominated by galaxies around $L_*$. 
%--- though in the
%simulations, too much of the luminosity and stellar mass expected in
%objects around $L_*$, is instead accumulated in the central cD.
In the significative  magnitude range, the shape of the predicted and 
observed LF is directly comparable. 
Yet, the simulated LF is steeper than the observed one, namely we 
underestimate
the relative number of bright galaxies, at least within the SW model here
shown.
We have found that increasing the feedback strength
results in decreasing the masses of all galaxies, and hence, in particular,
the number of bright galaxies. Alongside this, the LF in the Salpeter 
cases is broader in luminosity, due the lower stellar feedback
with respect to the AY IMF, allowing a larger accumulation of stellar 
mass in galaxies. 

%The LF in clusters can evolve due to two effects:
%passive luminosity evolution from the aging of the stellar populations
%in the galaxies, and mass evolution due to dynamical effects such as mergers, 
%tidal stripping and dynamical friction.
%Cluster galaxies are known to consist of old stellar populations, with the
%bulk of their star formation occurring at $z$$\gg$1. However,
%they may not have been completely assembled by $z\sim 1$, and they 
%may still undergo one or two significant mergers since then, or grow by gas 
%accretion (van Dokkum {\it et al.}, 1999). 

\begin{figure}
\centering \includegraphics[width=70mm]{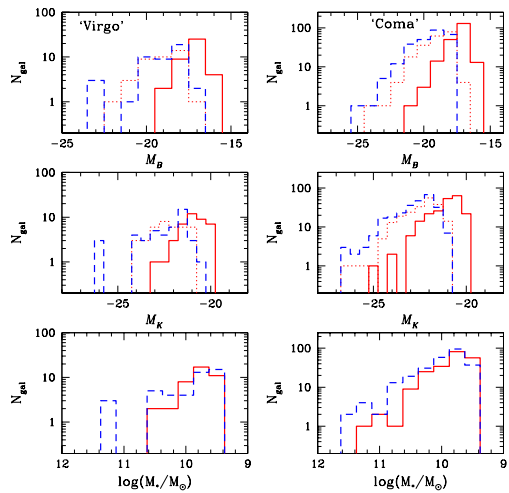}
%  \vspace*{174pt}
\caption{{\it Top panels:} Luminosity functions in absolute number of
galaxies per {\it B}-band magnitude bin for the standard (AY-SW) ``Virgo'' 
and ``Coma'' simulations, at $z=0$ ({\it solid 
lines}) and $z=1$ ({\it dashed lines}); the dotted lines represents
the expected $z=1$ LF if pure passive evolution of the stellar populations is
applied to the $z=0$ LF. {\it Middle panels:} Same, for the K-band luminosity
functions.
{\it Bottom panels:} Mass function of cluster
galaxies at $z=0$ and $z=1$.}
\label{fig:LFz}
\end{figure}

\begin{figure}
\centering \includegraphics[width=70mm]{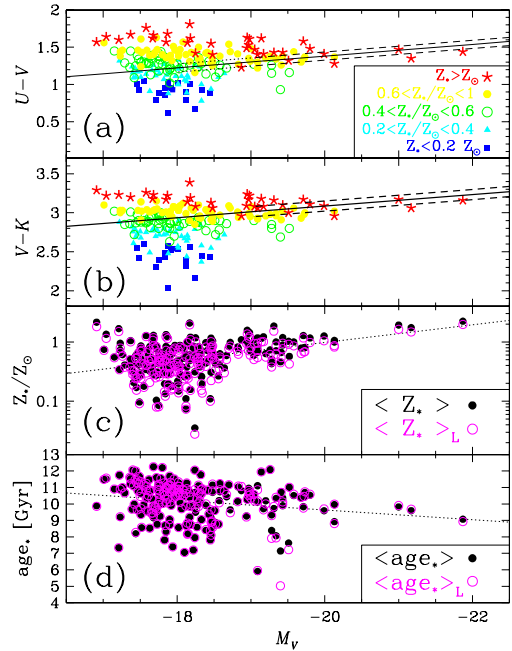}
%  \vspace*{174pt}
\caption{{\bf (a--b)} Colour--magnitude relation at $z$=0 for galaxies in the
simulated ``Coma'' cluster, compared to the observed relation with its scatter 
({\it solid} and {\it dashed} lines, from Bower {\it et~al.}\ 1992; 
{\it dotted} lines, from Terlevich {\it et~al.}\ 2001).
The extension of the dashed/dotted lines
indicates the magnitude range actually probed observationally, while the
solid line is an extrapolation of their fit to lower magnitudes.
{\bf (c)} Metallicity--luminosity relation for the ``Coma'' galaxies 
(full symbols for mass--averaged stellar metallicity, open symbols for 
luminosity--weighted metallicity); the dotted line is a linear fit.
{\bf (d)}: Age--luminosity relation for the ``Coma'' galaxies (mass--average
and luminosity--weighted stellar ages); the dotted line is a linear fit.}
\label{fig:redsequence}
\end{figure}

In Fig.~\ref{fig:LFz} we show the evolution of the B and K band LFs 
of our ``Virgo'' and ``Coma'' clusters. The $z=1$ LF is shifted to brighter 
magnitudes with respect to the distribution at $z=0$.
Both passive luminosity evolution and dynamical mass evolution 
are seen to play a role. In particular, 
since the bulk of the stars in our galaxies are formed at $z \gsim 2$ 
(see Fig.~\ref{fig:redsequence}, bottom panel),
luminosity dimming is an important effect.\\
The LF inferred at $z=1$ as expected from pure passive evolution of
the $z=0$ LF (dotted line
in Fig.~\ref{fig:LFz}) matches very closely the actual LF in the $z=1$ 
frame. 
%(apart from the absence of a few bright galaxies). 
This indicates that luminosity dimming of the stellar populations
drives most of the LF evolution and the fading to fainter 
magnitudes from $z=1$ to $z=0$ for the simulated galaxy population. 
%in line with some recent results (Kodama \& Bower, 2003).
In the brightest luminosity bins of our LF, some additional,
dynamical effect seems to be required. 
To assess mass evolution effects, we plot in the lower panels of
Fig.~\ref{fig:LFz} the mass function of cluster 
galaxies at $z=1$ and at $z=0$, which appear indeed very
similar.
``Over--merging'' onto the central cD in cosmological simulations can be 
responsible for depleting the number of galaxies at the bright end of the LF
(Springel {\it et~al.}\ 2001). Yet the results of a test simulation at 8 times 
higher mass and two times better force resolution we performed do allow us to
disregard this interpretation.

\subsection{The Red Sequence}
The light and stellar mass in clusters of galaxies is dominated by bright,
massive ellipticals, which are known
to form a tight colour-magnitude relation, or Red Sequence (Gladders {\it et~al.}\ 1998). In Fig.~\ref{fig:redsequence}ab, 
we compare the colour--magnitude relation for our ``Coma'' cluster galaxies,
excluding the cD, to two observed Red Sequences of Coma. 
There is overall agreement, although our Red Sequence is not as extended
as the observed one, and it also 
%(which reaches magnitudes brighter that $M_V=-23$), 
%due to the lack of simulated galaxies at the bright end of the luminosity 
%function already discussed. The simulated Red Sequence also
displays a larger scatter. 
%than the observed Coma Red Sequence, which remains 
%very narrow down to $M_V \sim -18$ (dashed and dotted lines).
The colour--magnitude relation is classically interpreted as a 
mass--metallicity relation. 
%(Kodama \& Arimoto 1997). In our simulations, 
The bulk of the stars in the simulated galaxies are formed at $z\ga$2, 
hence our galaxies are essentially coeval (Fig.~\ref{fig:redsequence}d) 
and their colour--magnitude relation is mostly a metallicity effect, as
it can be seen from combining these results with the 
metallicity--luminosity relation 
displayed in Fig.~\ref{fig:redsequence}c.

Moreover, the simulated Red Sequence appears to be a robust prediction
of our simulations, quite unaffected by the adopted physical prescriptions 
other than the chosen IMF. 
%The IMF sets the zero--point of the simulated 
%Red Sequence, via the typical stellar metallicity attainable in cluster
%galaxies; but the observed slope of the colour--magnitude relation
%is well reproduced by all simulations.

\bibliographystyle{plain}

\end{document}